\documentclass[aps,prl,twocolumn]{revtex4}

\usepackage{graphicx}

\begin{document}

%Title of paper
\title{Delayed Random Walks and Control}

\author{Tadaaki Hosaka}
\affiliation{Department of Computational Intelligence and Systems Science, Tokyo Institute of Technology, Yokohama, Japan 226-8502}
\thanks{current address: National Institute of Advanced Industrial Science and Technology,
Chuo-Dai2, 1-1-1, Umezono, Tsukuba, Ibaraki, 305-8568 Japan}
\author{Toru Ohira} 
\affiliation{Sony Computer Science Laboratories, Inc., Tokyo, Japan 141-0022}
\thanks{contact author: ohira@csl.sony.co.jp}

\begin{abstract}
 Issues of resonance that appear in non-standard random 
walk models are discussed. The first walk is called repulsive delayed random 
walk, which  
is described in the context of a stick balancing experiment. It will be shown that
a type of "resonant" effect takes place to keep the stability of the
fixed point better with tuned bias and delay.
We also briefly discuss the second model 
called  
sticky random walk, which is introduced to model string entanglement. 
Peculiar resonant effects with respect to these random walks are presented.  

\vspace{1em}
\noindent
(Published in {\it Flow Dynamics: The Second International Conference on Flow Dynamics} (M.Tokuyama and S. Maruyama eds.), AIP Conference Proceedings Vol. 832, pp. 487--491 (Melville, New York, 2006). )
\end{abstract}

\date{\today}
\maketitle

\section{Introduction}
A combination of non-linear dynamics and noise gives rise to the phenomena 
called stochastic  
resonance, which has been investigated actively 
\cite{longtin2,moss,collins,bulsara,gam}. The phenomena has been claimed 
to appear  
in a wide variety of things, such as climate change and neural information 
processing.  
The main theme of this paper is this type of phenomena in the context of  
non-standard random walks that we have proposed: repulsive 
\cite{ohira-hosaka04} and sticky. The former random walk was mainly  
derived from a stick balancing  
experiment\cite{cabrera-milton02,cabrera04,cabrera-etal04}, while the latter tries to 
model  
string entanglement. With both random walks, we observed rather 
unexpected phenomena  
that can be viewed as resonance. In the following, we describe each model 
and its  
associated behavior.

\section{Repulsive Delayed Random Walk} 
\subsection{Model} 

As a mathematical framework to investigate the systems with noise and 
delay, delayed random walk has been proposed and 
studied\cite{ohira-milton95,ohira97,ohira-yamane00}. This is a random 
walk whose transition probability depends on its position at a fixed time 
interval in the past. The focus has been placed on the model which has an 
attractive bias to a single point. This stable case has been applied to such 
processes like posture control \cite{collins-deluca94}. Analytically, the 
attractive delayed random walk model has shown such behaviors like an 
oscillatory correlation function with increasing delay. 

However, as the attractive model is not suitable to model the unstable 
situation we mentioned above, we discuss a delayed random walk which has 
a repulsive point. We can consider many different possibilities, but here we 
consider one-dimensional discrete time and step random walk with the 
origin as a repulsive point. Mathematically, we can define our model as 
follows. Let the position of the random walker at time step $t$ given by 
$X(t)$ and the fixed point set at the origin, $X=0$.
The delayed random walk is defined by the following conditional 
probabilities.
\begin{eqnarray}
P(X(t+1)= X(t)+1|X(t-\tau)>0) & = & p \\
P(X(t+1)= X(t)+1|X(t-\tau)=0) & = & {1\over2} \\
P(X(t+1)= X(t)+1|X(t-\tau)<0) & =& 1-p,
\end{eqnarray}
where $0 < p < 1$ and $\tau$ is the delay. With delay, the walker refers to 
its position in the past to decide on the bias of his next step. The attractive 
model is the case of $p < 0.5$, where the origin becomes attractive with no 
delay, $\tau=0$. On the other hand $p > 0.5$ gives the repulsive case which 
we shall discuss for the rest of this paper.

Though this appears to be a little change of definition from the attractive 
case, we observe a very different behavior from the attractive case. Most of 
all, as the walker escapes away from the origin, we do not have a stationary 
probability distribution. This makes analytical treatment of this repulsive 
model more difficult as compared to the attractive case, particularly with 
non-zero delay. Our investigation in this paper is done by computer 
simulation. The most notable feature of this model is that we can find an 
optimal combination of the bias  $p$ and $\tau$ where the random walker 
can be kept around the origin for a longest duration.

\subsection{Analysis and Simulation Results} 

As in the case of stick balance experiment, one of the main interests is how 
long the walker can be kept around the repulsive fixed point. We 
investigated this by focusing on an average first passage time
$L$ to reach a certain position (a limit point $\pm X^{*}, X^{*}>0$) away from the 
origin. In other words, we measured the average time for the walker starting 
from the origin to reach the limit point for the first time as we changed 
parameters in the model. The longer average first passage time indicates 
slower diffusion, which corresponds to the situation of longer stick balancing. 

For the case of zero delay with the bias $p$, we can find an analytical result 
for this average first passage time $L$ to reach the limit point $\pm X^{*}$ as
\begin{equation}
L = 2 \left({q\over{q-p}} \right)
         \left({{1- \left({q \over p} \right)^{X^*}}\over{1-{q \over p}}} 
\right)
         +{X^* \over {p-q}}, \quad (p \neq 0.5),
\end{equation}
where we have set $q \equiv 1-p$. For the case of simple (symmetric) 
random walk with
$p=q=0.5$, this result reduces to an even simpler form.
\begin{equation}
L = (X^*)^2.
\end{equation}
 
For the case of non-zero delay, such analytical result is yet to be obtained 
and computer simulation is used. We considered an ensemble of 10000 
walkers. The initial condition is set so that the walker performs a normal 
random walk with no bias $p=0.5$ for the duration of $t=(-\tau,0)$. The 
walker's position at $t=0$ is
set as the origin $X=0$. The limit point is set at $\pm X^{*}$. We measure 
the number of steps for each walker to go from the origin to $\pm X^{*}$ and 
average them. We performed computer simulations for various bias  
$p$ and delay $\tau$. 

Some sample results are shown in Figure 1. The most notable features of 
these graphs are the peaks in the graph, indicating that the slowest diffusion 
appears at certain optimal values of $\tau$ given bias $p$.
In other words, the walker is most stabilized around the origin with 
appropriate non-zero delay.
This is rather unexpected result contrary to the normal notion associated 
with effects of feedback delay,
where longer delay increasingly de-stabilize systems. Here, appropriate 
combination of bias and delay time is inducing more
stability.
\vspace{2em}

\begin{figure}[htbp]
\includegraphics[width=.4\textwidth]{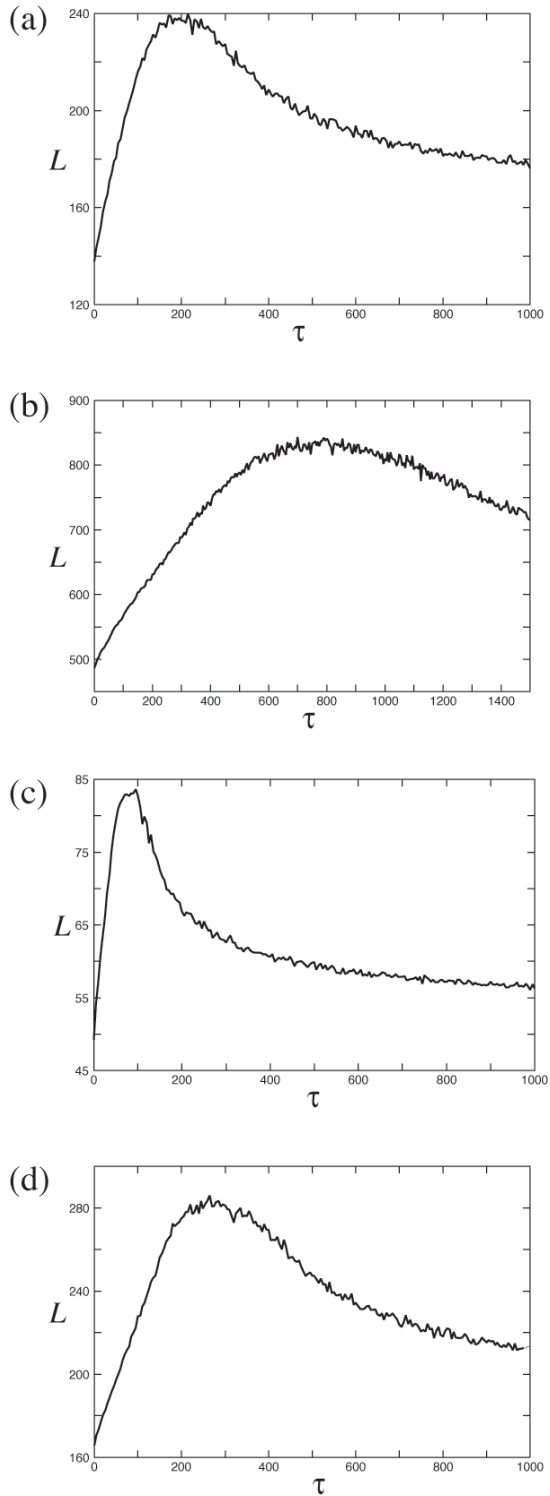}
\caption{Average first passage time $L$ as we change $\tau$. The value of 
parameters
$(p, {X^{*}})$ are (a) $(0.6, 30)$, (b) $(0.6, 100)$, (c) $(0.8, 30)$, and (d) $(0.8, 
100)$. }
\end{figure}
\clearpage

In order to gain more insight into this phenomenon, we look for an 
approximate analytical expression, which is found to be given by the 
following expression 
\begin{equation}
L(\tau) =(1 + \alpha \tau_n {e^{ -\beta \tau_n }})L(\tau=0).
\end{equation}
Here $\alpha$ and $\beta$ are parameters and
${\tau_n}$ is a normalized delay given as follows.
\begin{equation}
{\tau_n}\equiv \tau { p-q \over {X^{*}} }.
\end{equation}
This normalization uses a characteristic time dividing the distance 
$X^{*}$ by an average velocity of the walker
 $p-q$. Hence ${\tau_n}$ is a non-dimensionalized parameter as well.

\begin{figure}[htbp]
\includegraphics[width=.4\textwidth]{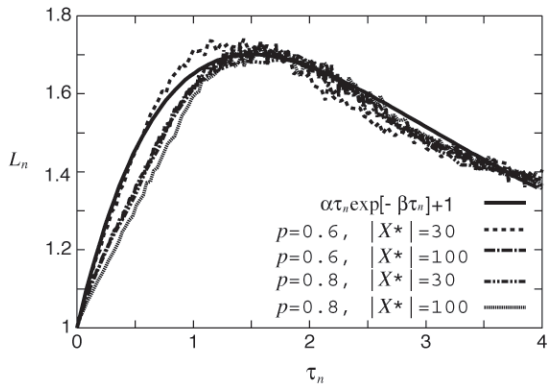}
\caption{Normalized average first passage time $L_n\equiv {L(\tau)\over L(\tau=0)}$ as we change 
normalized delay $\tau_n$. The parameter sets $(p, {X^{*}})$ plotted 
are $(0.6, 30)$,  $(0.6, 100)$, $(0.8, 30)$,  and $(0.8, 100)$ }
\end{figure}

Figure 2 shows this analytical approximation and the result of computer 
simulation. We see that the curves for various  $(p, {X^{*}})$ overlaps quite 
well with the analytical curve with appropriately chosen parameters of 
$\alpha=1.27$ and $\beta=0.67$. We also notice that the peak height is 
approximately 1.7 times the average first passage time of zero delay case.

\subsection{Delayed Stochastic Control} 
 These theoretical results imply that systems can reach a better balancing  
performance if an appropriate amount of fluctuation is added given the 
feedback or  
reaction delay. We have termed this type of control, which is different from 
standard  
feedback or predictive ones, as delayed stochastic control. We performed  
the following experiment to gain some  
insight into the existence or utilization of this control scheme. We asked the 
subjects to sit on a chair and balance a  
stick, as in the previous stick balancing experiment. But, this time, the 
subjects  
were allowed to move their bodies, not just their arms, as they tried to 
balance the stick. One way to do this is to  
hold an object with the other hand and move it (Figure 3). Another way is to 
move  
their legs. We measured the time for which they could keep the sticks 
balanced, and  
compared it with the normal non-movement situations. Out of the six 
subjects we tested,  
three subjects showed notable improvement in balancing by  
reaching their own optimal level of movement (Figure 4).  
\begin{figure}[htbp]
\includegraphics[width=.35\textwidth]{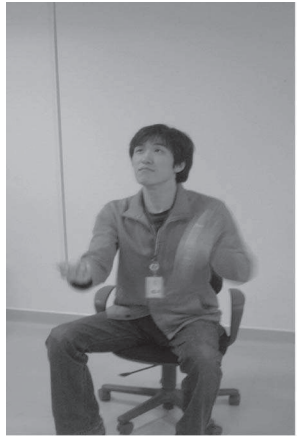} 
\caption{Picture of a subject balancing a stick on one hand while moving an 
object  
in the other. } 
\end{figure} 
\begin{figure} [htbp]
\includegraphics[width=.36\textwidth]{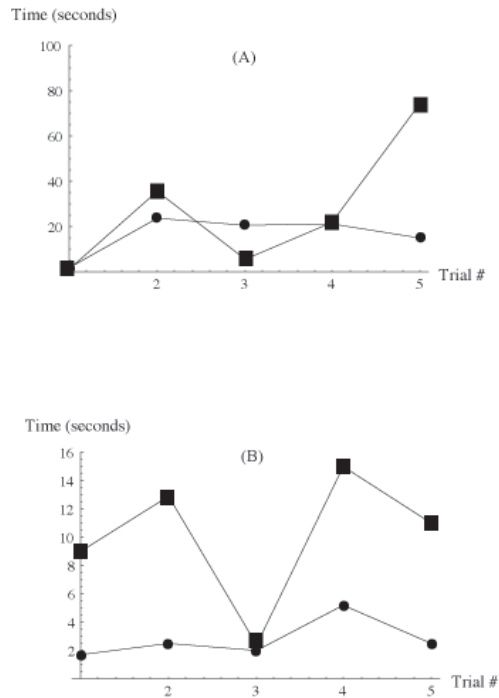} 
\caption{(A) Example of improvement on balancing tasks with (square) and 
without (dot) moving an object.  
The subject was given 5 trials without previous practice. By the 5th trial, the 
improvement was  
significant. (B) Another subject practiced for a few hours. Here, again 
improvement with 
moving the object was evident.} 
\end{figure} 
\clearpage
 
Some practice was needed for  
these subjects to reach this better performance. We believe that  
the subjects were tuning the appropriate level of fluctuation given their 
reaction times and  
prediction accuracy. Even though more thorough data needs to be collected, 
these results  
may be one supporting example of delayed stochastic control.

\section{Sticky Random Walk} 
\subsection{Model} 
 
Entangled strings is something we commonly observe. For example, wires for 
electrical appliances or communication network cords sometimes require  
us to disentangle them. We describe here a concept of sticky random walk we 
used  
to gain some insight into this phenomenon. The model is simple.  
The strings are represented by the trajectory of a random walker. This 
random walker  
leaves sticks or marks at certain time intervals. Therefore, a string is  
represented by this trajectory with these marks on it. By sending out 
multiple sticky  
random walkers, we obtained multiple sticky strings. Furthermore, a string 
is considered  
as entangled with another when these marks overlap at the same site in 
space,  
and not when they are simply crossed. Thus, the string is considered more 
sticky when there  
are more marks on it. 

We tested a situation having multiple sticky strings in  
a bounded two-dimensional square grid by sending out sticky random walks 
in this space.  
These random walks are discrete time, discrete space walks moving one step 
to its  
neighboring grid points. They are bounded by the edge of the square grid. We 
then  
pick one string randomly and count the number of strings either directly or 
indirectly   
entangled to that string. Indirect entanglement indicates that two strings 
are entangled  
through others, i.e., two strings can reach each other by following the chain  
of directly entangled strings. We performed simulation experiments with 
various conditions on the number of strings, the number  
of marks on each string, the length of each string, and the size of the 
two-dimensional square  
grid. In particular, we asked the question,  
if we compare the situation of having more strings with fewer marks and 
that of  
having fewer strings with more marks, while keeping the total number of 
marks in the  
space constant, which situation gives rise to more entanglement?  
 
\subsection{Simulation Results} 
 
We kept the total number of sticky marks $R$ and the length of each string 
$L$ as  
fixed, and we varied the number of strings $S$ and marks on each string 
$M$ so that $R =  
M \times S$. The number of entangled strings was measured both in 
numbers $E$ and  
in ratio $e = {E \over S}$. Each part of the data is an average over 100 trials, 
with various  
space for $N$ by $N$ square grid. The representative results are shown in 
Figure 5. We  
found that an optimal combination of $S$ and $M$ exists. It is given as the  
highest peak in these graphs. This means that these strings are most 
entangled when  
the level of stickiness and the number of strings are optimally tuned. Even 
more unexpectedly,  
this optimal combination is independent of the space size $N$ for the ratio 
$e$. When  
$N$ is sufficiently large, it is also independent with respect to $E$ as well. 
Though  
it differs from the standard form of stochastic resonance, randomness in the 
motion of  
the walkers plays a role in bringing about this resonant behavior. Whether 
or not this behavior  
can happen in a real situation requires experimental tests. 
 
\begin{figure}
\includegraphics[width=.4\textwidth]{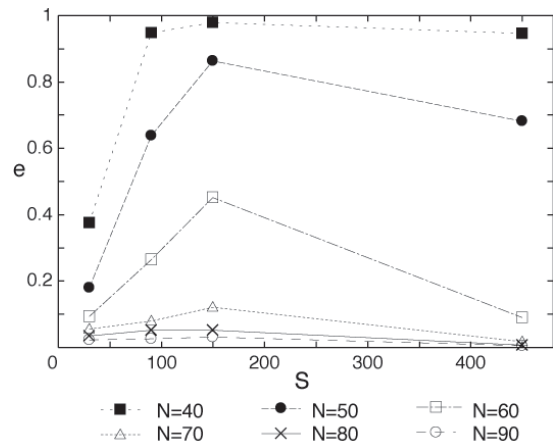} 
\caption{Average ratio, $e$, of entangled strings $e$ as  
number of marks on each string is changed. The length of each string is set 
at  
$L=60$, and the total number of marks is set at $R=1800$. Each line 
corresponds to  
a square lattice size $N$. } 
\end{figure} 
 
\section{Discussion}

We discussed two non-standard models of stochastic resonance. 
As a related subject, a binary bit model that shows resonance with noise and delay
are proposed and studied \cite{ohira-sato,tsimring}. 
This phenomena was observed in an experiment with solid state laser
\cite{masoller}.

Our  
investigations here with respect to these resonance with random walks are still in the beginning stages. However, they already 
produced  
quite unexpected results. 
Further analysis as well as application with real  
systems could lead to some additional interesting insights.

%%%%%%%%%%%%%%%%%%%%%%%%%%%%%%%%%%

\subsubsection*{Acknowledgments}
 We thank Dr. Juan Luis Cabrera and  Prof. John G. Milton for their  
insightful  
comments. 
TH was a Research Fellow of the Japan Society for the Promotion of Science 
(JSPS). 
We acknowledge a support from Grant-in-Aid No. 164453 from the JSPS.

\end{document}